\documentstyle[12pt]{article}

\begin{document}

\begin{titlepage}
\begin{flushright}
DFUPG 110/95
\end{flushright}

\vspace{1cm}
\begin{centering}

{\LARGE Mesh Currents and Josephson Junction Arrays}
\\
\vspace{0.7cm}
{\large Carlo Lucheroni$^*$}
\vspace{0.5cm}
\\
{Dipartimento di Fisica, Universit\`a di Perugia, 06100 Perugia, Italy}
\\
\vspace{0.5cm}
{November, 1995$^{**}$}\\
\end{centering}
\vspace{.25cm}

\begin{abstract}
A simple but accurate mesh current analysis is performed on a XY model and
on a SIMF model to derive the equations for a Josephson junction array. The
equations obtained here turn out to be different from other equations
already existing in the literature. Moreover, it is shown that the two
models come from a unique hidden structure.   
\end{abstract}

\vspace{1.3cm}

\begin{centering}
\footnotesize {$^*$e-mail address: lucheroni@vaxpg.pg.infn.it}\\
\end{centering}
\begin{centering}
\footnotesize {$^{**}$revised on May, 1996}\\
\end{centering}

\end{titlepage}


\section{Introduction}

This is an analysis of a planar electric network made of Josephson junctions
connected by superconducting wires, biased by external currents in the
presence of a perpendicular external static magnetic field. Such a circuit
is usually named Josephson array and it is interesting for technological and
theoretical reasons. It is modeled by a system of coupled nonlinear
dynamical equations, and its physics is based on the behavior of
supercurrents interacting directly each other and also by induced magnetic
flux. When there is no external bias, starting from any initial condition
the supercurrents reach an equilibrium state and then circulate forever
without dissipation, sometimes trapping flux in the network meshes and
accomodating themselves in characteristic patterns which also depend on the
external magnetic field. In the presence of bias these currents can
rearrange and adjust to the new external conditions, up to a certain bias
limit. If this limit is exceeded by the bias, an out of equilibrium
dissipative state is reached and flux patterns can move through. The circuit
then emits a radiation originated in its elements: if synchronization sets
up, this radiation can be constructively coherent and much more energetic
than the radiation from a single junction. The exploitation of this
synchronization mechanism is the technical reason for the interest for such
a system, which allows making usable the feeble but very high frequency and
easily tunable radiation emitted from a single Josephson junction.

The theoretical interest is in the understanding of the complex dynamics
underlying the dance of the flux patterns and the action of the magnetic
field on both static and dynamic state. To achieve this result, the
attention was focused on the choice of the model to be used to describe this
system. As pointed out in Ref.\cite{fila2}, two seemingly different classes
of models have been selected:\ the first one is referred to as {\it XY} {\it %
model} and the second one can be referred to as {\it self-induced magnetic
field} (SIMF) model. The XY model derives from field theory \cite{halsey}
and indirectly explains many array phenomena, but is limited to systems in
wich array inductance is negligible. The SIMF model takes into account the
induced flux interaction and then should be valid in every situation,
including the case when the inductance goes smoothly to zero. Exactly at
zero inductance it should merge with the XY model, but this cannot be proved
because of the presence of a parameter singularity. The SIMF model is more
difficult to be studied analytically than the XY model, but in the form
studied in Ref.\cite{naka} it has a rare direct experimental evidence \cite
{chris},\cite{fuji}. It is anyway thought to be fundamental \cite{dominguez}
to explain some peculiar features in giant Shapiro steps and other related
phenomena. In Ref.\cite{mit} a compact formalization of the SIMF model was
given as an explicit dynamical system. In Ref.\cite{basler} an attempt was
made to study analitically the parameter region where the two models should
merge, but the singularity in the inductance parameter makes the analysis
difficult. A numerical investigation of this parameter region was made in
Ref.\cite{dominguez}, but a numerical simulation drawback is the difficulty
to get a global understanding of how the transition works, beeing limited
only to finite parameter ranges. Ref.\cite{dominguez} is also a source for
more references.

In this work two main conclusions about these models are reached. A simple
but careful derivation of the equations leads to systems which are different
from those existing in the previous literature. Moreover, XY and SIMF models
derive from the same set of equations and the study of the transition region
is easier than it was thought before. The plan of the paper is to work out a
simple reference example in Section 2 and to derive in a general way the
equations for a XY and a SIMF rectangular network in Section 3. Then in
Section 4 the source of the difference will be briefly discussed. In Section
5 the system from which both models derive will be shown.

\section{Four circuits as an example}

\input{epsf.tex} 
\begin{figure}[ht]
\begin{center}
\fbox{
\epsfxsize=7.truecm \epsffile{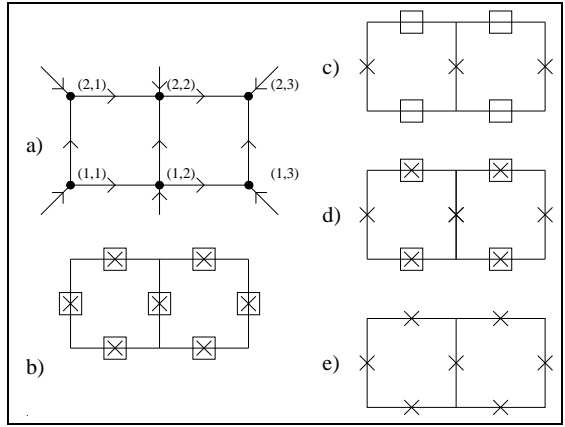}
}
\end{center}
\caption{Example circuits and notation}
\label{fig1}
\end{figure}

In principle, a Josephson array can be built laying a rectangular grid of
superconducting stripes on an insulating substrate so that cells surrounded
by superconducting wires are formed. A rectangular array has $N_r\times N_c$
cells, where $N_r$ indicates the number of cells in one row and $N_c$ the
number of cells in one column. In the examples of Fig.\ref{fig1} the cells
are square and each circuit consists in two cells. In circuit $a)\,$
conventions are displayed: nodes are labeled by the two row and column
indices $(r,c)$ and reference directions for the currents are chosen.
Orientated currents will be called $i_{v,r,c}$ and orientated gauge
invariant phases will be called $\varphi _{v,r,c}$ , where $r$ and $c$ refer
to the node where they originate from, index $v=0$ is for horizontal
branches and $v=1$ for vertical branches. Meshes are labeled only with the
two indices $(r, c)$ of their bottom left hand corner and orientated
clockwise. External bias currents are called $I_{r,c}$ where the indices $r$
and $c$ refer to the node where the current is injected. The Josephson
junctions considered here are proximity effect junctions, which can be
obtained cutting the branch wire in the center, and they are indicated by
crosses. Small boxes will be explained later. The current $i$ flowing along
a branch containing a junction is described by the resistively shunted (RSJ)
model \cite{likha} as 
\begin{equation}
\label{rsj}i=i_n+i_s=\frac 1R\frac d{dt}\varphi \,+\,i_c\sin \,\varphi 
\end{equation}
in terms of the superconductor order parameter phase difference $\varphi $
between the edges of the junction: $R$ is a phenomenological resistence and $%
i_c$ is the maximum supercurrent that can flow in the junction. The first $%
i_n$ term arises from the voltage $V=\frac \hbar {2e}\stackrel{.}{\varphi \;}
$which sets in due to a.c. Josephson effect when $\varphi $ varies in time,
the second $i_s$ term describes the $V=0$ supercurrent that flows by Cooper
pairs tunneling. To treat $\varphi $ as a real variable and not as a
distribution, an assumption of very low temperature conditions is made here.
Three sets of equations will now be written \cite{likha} for each of the
circuits $b)$, $c)$, $d)$ and $e)$ in Fig.\ref{fig1}. Since circuits $b)$, $%
d)$ and $e)$ have junctions in every branch they will be named {\it pure}
Josephson networks while circuit $c)$ will be named {\it hybrid}, lacking
junctions in some branches.

When currents circulate in an array, each wire induces a magnetic field
proportional to the current $i$ flowing in the wire itself. If this magnetic
field is not too strong, each cell is threaded only by the {\it self-induced
magnetic flux} $\phi _{r,c}$ generated by the closed current formed by the
currents $i_{v,r,c}$ circulating in the wires along the borders of the mesh $%
(r,c)$, plus a possible {\it external magnetic flux} $\phi _{r,c}^{ext}$. If 
$\phi _{r,c}^{tot}$ is the total flux, 
\begin{equation}
\phi _{r,c}^{tot}=\phi _{r,c}^{ext}-\phi _{r,c} 
\end{equation}
For mesh $(1,1)$ and $(1,2)$ in circuit $b)$ induced fluxes are 
\begin{equation}
\label{indu1}\left\{ 
\begin{array}{c}
\phi _{1,1}=L(i_{1,1,1}+i_{0,2,1}-i_{1,1,2}-i_{0,1,1}) \\ 
\phi _{1,2}=L(i_{1,1,2}+i_{0,2,2}-i_{1,1,3}-i_{0,1,2}) 
\end{array}
\right. 
\end{equation}
where $L$ is the cell self-inductance. In the circuits in Fig.\ref{fig1}
some branches have small boxes attached to indicate generation of flux,
while branches without boxes generate negligible flux. If a junction is
present in a wire, it limits to $i_c$ the current that can flow in it; if
the junction critical current $i_c$ is very low, the wire contributes to
induced flux only with the $i_n$ normal term, which at equilibrium is null.
In fact, junctions not enclosed in boxes have critical currents $i_c^{\prime
}$ negligible in comparison to others.

In this aspect hybrid array $c)$ and pure array $d)$ are analogous: in $c)$
vertical branches contributions to the flux are negligible in comparison to
contributions from unbroken horizontal wires. In $d)$ junctions with very
low $i_c^{\prime }<i_c$ are present in vertical branches and their induced
flux is negligible in comparison to that from horizontal branches. Then for
circuits $c)$ and $d)$

\begin{equation}
\label{indu2}\left\{ 
\begin{array}{c}
\phi _{1,1}=L(i_{0,2,1}-i_{0,1,1}) \\ 
\phi _{1,2}=L(i_{0,2,2}-i_{0,1,2}) 
\end{array}
\right. 
\end{equation}
while in circuit $e)$ 
\begin{equation}
\label{indu3} \phi _{1,1}=\phi _{1,2}=0 
\end{equation}
which could be naively seen as a $L\rightarrow 0$ limit of Eq.(\ref{indu1}).
Eq.(\ref{indu1}), (\ref{indu2}) and (\ref{indu3}) are different instances of
the first set of equations. Circuits $b)$, $c)$ and $d)$ are SIMF models
while circuit $e)$ is a prototypical XY model.

The second set of equations comes from the principle that the directed sum
of gauge invariant phase differences around a mesh is related to the induced
flux threading the mesh \cite{likha}: 
\begin{equation}
\label{fluxoid}\left\{ 
\begin{array}{c}
\phi _{1,1}=-(\varphi _{1,1,1}+\varphi _{0,2,1}-\varphi _{1,1,2}-\varphi
_{0,1,1})+f_{1,1} \\ 
\phi _{1,2}=-(\varphi _{1,1,2}+\varphi _{0,2,2}-\varphi _{1,1,3}-\varphi
_{0,1,2})+f_{1,2} 
\end{array}
\right. 
\end{equation}
where $f_{r,c}=2\pi \,n_{r,c}+\phi _{r,c}^{ext}$ are called for historical
reasons {\it frustrations} and $n_{r,c}$ are integers. In $\phi _{1,1}$ ,$%
\phi _{1,2}$, $\phi _{r,c}^{ext}$ a coefficient $\frac{4\pi e}h$ has been
absorbed, where $e$ is the electron charge and $h$ the Plank constant. The
sums $(\varphi _{1,1,1}+\varphi _{0,2,1}-\varphi _{1,1,2}-\varphi
_{0,1,1}+\phi _{1,1}-\phi _{r,c}^{ext})$ and $(\varphi _{1,1,2}+\varphi
_{0,2,2}-\varphi _{1,1,3}-\varphi _{0,1,2}+\phi _{1,2}$ $-\phi _{r,c}^{ext})$
are named {\it fluxoids} and Eq.(\ref{fluxoid}) states that fluxoids are
equal to $2\pi $ times an arbitrary integer $n$.

The third set of equations implements Kirchhoff current conservation law for
nodes in an electric network. The oriented currents sum to zero for each
node as 
\begin{equation}
\label{kirch} 
\begin{array}{cc}
\left\{ 
\begin{array}{l}
-i_{1,1,1}-i_{0,1,1}+I_{1,1}=0 \\ 
i_{1,1,1}-i_{0,2,1}+I_{2,1}=0 \\ 
i_{0,2,1}+i_{1,1,2}-i_{0,2,2}+I_{2,2}=0 
\end{array}
\right. , & \left\{ 
\begin{array}{l}
i_{0,2,2}+i_{1,1,3}+I_{2,3}=0 \\ 
-i_{1,1,3}+i_{0,1,2}+I_{1,3}=0 \\ 
-i_{0,1,2}-i_{1,1,2}+i_{0,1,1}+I_{1,2}=0 
\end{array}
\right. 
\end{array}
\end{equation}
which is valid for every circuit.

From these three sets of equations a nonlinear differential system for each
circuit is going to be derived, where only one phase derivative will appear
explicitly for each equation. Such a system is called {\it explicit}
dynamical system. For simplicity, from now until differently stated, it will
be assumed that external fluxes $\phi _{r,c}^{ext}$ and integers $n_{r,c}$
are zero, and $R=1$. A bias current $\gamma $ enters from the upper nodes
and leaves the circuit from the lower nodes as $-\gamma $: then $%
I_{2,1}=I_{2,2}=I_{2,3}=\gamma $ and $I_{1,1}=I_{1,2}=I_{1,3}=-\gamma $.
System (\ref{kirch}) consists of $6$ linear equations in $7$ unknowns, but
only $5$ equations are linearly independent. Two free parameters $I_{1,1}^a$
and $I_{1,2}^a$ are introduced, named {\it mesh currents}, to solve it: 
\begin{equation}
\label{sol1} 
\begin{array}{cc}
\left\{ 
\begin{array}{l}
i_{1,1,1}=I_{1,1}^a-\gamma \\ 
i_{1,1,2}=-I_{1,1}^a+I_{1,2}^a-\gamma \\ 
i_{1,1,3}=-I_{1,2}^a-\gamma 
\end{array}
\right. , & \left\{ 
\begin{array}{l}
i_{0,1,1}=-I_{1,1}^a \\ 
i_{0,1,2}=-I_{1,2}^a \\ 
i_{0,2,1}=I_{1,1}^a \\ 
i_{0,2,2}=I_{1,2}^a 
\end{array}
\right. 
\end{array}
\end{equation}
Direct substitution in system (\ref{kirch}) works as a proof. Induced fluxes
in systems (\ref{indu1}) and (\ref{indu2}) can be expressed in terms of Eq.(%
\ref{sol1}) too: 
\begin{equation}
\label{mu} 
\begin{array}{cc}
(\ref{indu1})\rightarrow \left\{ 
\begin{array}{l}
\phi _{1,1}=L(4I_{1,1}^a-\;I_{1,2}^a) \\ 
\phi _{1,2}=L(-\;I_{1,1}^a+4I_{1,2}^a) 
\end{array}
\right. , & (\ref{indu2})\rightarrow \left\{ 
\begin{array}{l}
\phi _{1,1}=2LI_{1,1}^a \\ 
\phi _{1,2}=2LI_{1,2}^a 
\end{array}
\right. 
\end{array}
\end{equation}
An immediate remark has to be made here: the system from (\ref{indu1})
relates the self-induced flux $\phi _{r,c}$ with all the possible mesh
currents whereas the system from (\ref{indu2}) relates the flux $\phi _{r,c}$
only to the mesh current $I_{r,c}^a$ bearing the same indices $(r,c)$. These
systems can be inverted to give 
\begin{equation}
\label{mi} 
\begin{array}{cc}
(\ref{indu1})\rightarrow \left\{ 
\begin{array}{l}
I_{1,1}^a=\frac 1{15}\frac 1L\left( 4\phi _{1,1}+\;\phi _{1,2}\right) \\ 
I_{1,2}^a=\frac 1{15}\frac 1L\left( \phi _{1,1}+\;4\phi _{1,2}\right) 
\end{array}
\right. , & (\ref{indu2})\rightarrow \left\{ 
\begin{array}{l}
I_{1,1}^a=\frac 12\frac 1L\phi _{1,1} \\ 
I_{1,2}^a=\frac 12\frac 1L\phi _{1,2} 
\end{array}
\right. 
\end{array}
\end{equation}
and such mesh currents can be inserted in Eq.(\ref{sol1}), which now relates
branch currents $i_{v,r,c}$ to self-induced fluxes $\phi _{r,c}$.

The differential equations for the circuit $b)$ can now be written
substituting in system (\ref{sol1}) Eq.(\ref{rsj}) for the l.h.s. and the
first block of Eq.(\ref{mi}) for the r.h.s., where $\phi _{r,c}$ are
obtained in turn from Eq.(\ref{fluxoid}%
$$
b)\negthinspace \,\negthinspace \left\{ 
\begin{array}{l}
\stackrel{.}{\varphi }_{0,1,1}+i_c\sin (\varphi _{0,1,1})= \\ \frac{-1}{15L}%
\negthinspace \left( 4\varphi _{0,1,1}+\varphi _{0,1,2}-4\varphi
_{1,1,1}+3\varphi _{1,1,2}+\varphi _{1,1,3}-4\varphi _{0,2,1}-\varphi
_{0,2,2}\right) \\ \stackrel{.}{\varphi }_{0,1,2}+i_c\sin (\varphi
_{0,1,2})= \\ \frac{-1}{15L}\negthinspace \left( \varphi _{0,1,1}+4\varphi
_{0,1,2}-\varphi _{1,1,1}-\varphi _{1,1,2}+4\varphi _{1,1,3}-\varphi
_{0,2,1}-4\varphi _{0,2,2}\right) \\ \stackrel{.}{\varphi }_{1,1,1}+i_c\sin
(\varphi _{1,1,1})= \\ \frac{-1}{15L}\negthinspace \left( -4\varphi
_{0,1,1}-\varphi _{0,1,2}+4\varphi _{1,1,1}-3\varphi _{1,1,2}-\varphi
_{1,1,3}+4\varphi _{0,2,1}+\varphi _{0,2,2}\right) -\gamma \\ \stackrel{.}{%
\varphi }_{1,1,2}+i_c\sin (\varphi _{1,1,2})= \\ \frac{-1}{15L}\negthinspace %
\left( 3\varphi _{0,1,1}-3\varphi _{0,1,2}-3\varphi _{1,1,1}+6\varphi
_{1,1,2}-3\varphi _{1,1,3}-3\varphi _{0,2,1}+3\varphi _{0,2,2}\right)
-\gamma \\ \stackrel{.}{\varphi }_{1,1,3}+i_c\sin (\varphi _{1,1,3})= \\ 
\frac{-1}{15L}\negthinspace \left( \varphi _{0,1,1}+4\varphi
_{0,1,2}-\varphi _{1,1,1}-3\varphi _{1,1,2}+4\varphi _{1,1,3}-\varphi
_{0,2,1}-4\varphi _{0,2,2}\right) -\gamma \\ \stackrel{.}{\varphi }%
_{0,2,1}+i_c\sin (\varphi _{0,2,1})= \\ \frac{-1}{15L}\negthinspace \left(
-4\varphi _{0,1,1}-\varphi _{0,1,2}+4\varphi _{1,1,1}-3\varphi
_{1,1,2}-\varphi _{1,1,3}+4\varphi _{0,2,1}+\varphi _{0,2,2}\right) \\ 
\stackrel{.}{\varphi }_{0,2,2}+i_c\sin (\varphi _{0,2,2})= \\ \frac{-1}{15L}%
\negthinspace \left( -\varphi _{0,1,1}-4\varphi _{0,1,2}+\varphi
_{1,1,1}+3\varphi _{1,1,2}-4\varphi _{1,1,3}+\varphi _{0,2,1}+4\varphi
_{0,2,2}\right) 
\end{array}
\right. 
$$
This result is remarkably different in the form of the interaction among
phases from those in Ref.\cite{mit} and \cite{fila}. Here in {\it each}
equation are present {\it all} the phases of the system, while there in each
equation is present only a subset of the whole set of phases.

The differential equations for the hybrid circuit $c)$ are obtained in a
slightly different way, because orizontal branch currents $i_{0,r,c}$ are 
{\it not} described by Eq.(\ref{rsj}). In this case branches are considered
as containing inductances and not as simple shorts. Substituting $\phi
_{r,c} $ from Eq.(\ref{indu2}) in the second block of Eq.(\ref{mi}) gives $%
I_{r,c}^a $ in terms of $i_{v,r,c}$. In turn $I_{r,c}^a$ are inserted in
system (\ref{sol1}). A set of $4$ relations $i_{0,1,1}=-\frac
12(i_{0,2,1}-i_{0,1,1})$, $i_{0,2,1}=\frac 12(i_{0,2,1}-i_{0,1,1})$, $%
i_{0,1,2}=-\frac 12(i_{0,2,2}-i_{0,1,2})$ and $i_{0,2,2}=\frac
12(i_{0,2,2}-i_{0,1,2})$ is obtained and it is satisfied iff $%
i_{0,1,1}=-i_{0,2,1}$ and $i_{0,1,2}=-i_{0,2,2}$. This implies that the $6$
equations in system (\ref{kirch}) are reduced to the $3$ equations 
\begin{equation}
\left\{ 
\begin{array}{l}
i_{1,1,1}=i_{0,2,1}-\gamma \\ 
i_{1,1,2}=i_{0,1,1}-\gamma \\ 
i_{1,1,3}=i_{0,2,2}-\gamma 
\end{array}
\right. 
\end{equation}
where Eq.(\ref{rsj}) is to be applied only to the l.h.s. terms. Eq.(\ref
{fluxoid}) for circuit $c)$ has to be modified in $\phi _{1,1}=-(\varphi
_{1,1,1}-\varphi _{1,1,2})$ and $\phi _{1,2}=-(\varphi _{1,1,2}-\varphi
_{1,1,3})$, because only two junctions are present in each cell. Using this
relation gives 
$$
c)\;\left\{ 
\begin{array}{l}
\stackrel{.}{\varphi }_{1,1,1}+i_c\sin (\varphi _{1,1,1})=-\frac
1{2L}(\varphi _{1,1,1}-\varphi _{1,1,2})\ -\gamma \\ \stackrel{.}{\varphi }%
_{1,1,2}+i_c\sin (\varphi _{1,1,2})=\frac 1{2L}(\varphi _{1,1,1}-2\varphi
_{1,1,2}+\varphi _{1,1,3})-\gamma \\ \stackrel{.}{\varphi }_{1,1,3}+i_c\sin
(\varphi _{1,1,3})=\frac 1{2L}(\varphi _{1,1,2}-\varphi _{1,1,3})-\gamma 
\end{array}
\right. 
$$
This result is the same as that in Ref.\cite{naka}, where it was derived for
an infinite stripe of cells. For this circuit, in contrast to the circuit $%
b) $, in each equation is present only a subset of all possible phases. This
is clearly due to the fact that in every mesh of this circuit the
self-induced flux $\phi _{r,c}$ is proportional only to the $I_{r,c}^a$ with
the same indices $(r,c)$.

The differential equations for the pure array $d)$ are derived in the same
way as for circuit $b)$, substituting in system (\ref{sol1}) Eq.(\ref{rsj})
for the l.h.s. and the second block of Eq.(\ref{mi}) for the r.h.s., where
in turn $\phi _{r,c}$ are obtained from Eq.(\ref{fluxoid}):%
$$
d)\,\left\{ 
\begin{array}{l}
\stackrel{.}{\varphi }_{0,1,1}+i_c\sin (\varphi _{0,1,1})= \\ -\frac 1L\frac
12\left( \varphi _{0,1,1}-\varphi _{1,1,1}+\varphi _{1,1,2}-\varphi
_{0,2,1}\right) \\ 
\stackrel{.}{\varphi }_{0,1,2}+i_c\sin (\varphi _{0,1,2})= \\ -\frac 1L\frac
12\left( \varphi _{0,1,2}-\varphi _{1,1,2}+\varphi _{1,1,3}-\varphi
_{0,2,2}\right) \\ 
\stackrel{.}{\varphi }_{1,1,1}+i_c^{\prime }\sin (\varphi _{1,1,1})= \\ 
-\frac 1L\frac 12\left( -\varphi _{0,1,1}+\varphi _{1,1,1}-\varphi
_{1,1,2}+\varphi _{0,2,1}\right) -\gamma \\ 
\stackrel{.}{\varphi }_{1,1,2}+i_c^{\prime }\sin (\varphi _{1,1,2})= \\ 
-\frac 1L\frac 12\left( \varphi _{0,1,1}-\varphi _{0,1,2}-\varphi
_{1,1,1}+2\varphi _{1,1,2}-\varphi _{1,1,3}-\varphi _{0,2,1}+\varphi
_{0,2,2}\right) -\gamma \\ 
\stackrel{.}{\varphi }_{1,1,3}+i_c^{\prime }\sin (\varphi _{1,1,3})= \\ 
-\frac 1L\frac 12\left( \varphi _{0,1,2}-\varphi _{1,1,2}+\varphi
_{1,1,3}-\varphi _{0,2,2}\right) -\gamma \\ 
\stackrel{.}{\varphi }_{0,2,1}+i_c\sin (\varphi _{0,2,1})= \\ -\frac 1L\frac
12\left( -\varphi _{0,1,1}+\varphi _{1,1,1}-\varphi _{1,1,2}+\varphi
_{0,2,1}\right) \\ 
\stackrel{.}{\varphi }_{0,2,2}+i_c\sin (\varphi _{0,2,2})= \\ -\frac 1L\frac
12\left( -\varphi _{0,1,2}+\varphi _{1,1,2}-\varphi _{1,1,3}+\varphi
_{0,2,2}\right) 
\end{array}
\right. 
$$
This system has the same polarization scheme as systems $b)$ and $c)$ and a
limited interaction among phases: only in the differential equation for
phase $\varphi _{1,1,2}$ all the other phases appear. This is again due to
the proportionality of mesh currents and induced fluxes. It seems having
never been discussed in the literature, even though it has an interaction
equal to that attributed to circuit{\it \ }$b)$ in Ref.\cite{mit}.

The differential equations for the last circuit $e)$ are obtained in another
different way. In this case induced fluxes are zero, so it is not possible
to use them to eliminate the mesh currents in system (\ref{sol1}). Anyway,
Eq.(\ref{fluxoid}) is still valid and can be differentiated. Inserting Eq.(%
\ref{rsj}) in system (\ref{sol1}) together with differentiated Eq.(\ref
{fluxoid}) gives a system that can be rewritten as%
$$
\begin{array}{cc}
\left\{ 
\begin{array}{l}
\stackrel{.}{\varphi }_{0,1,1}+\,i_c\sin (\varphi _{0,1,1})=-I_{1,1}^a \\ 
\stackrel{.}{\varphi }_{0,1,2}+\,i_c\sin (\varphi _{0,1,2})=-I_{1,2}^a \\ 
\stackrel{.}{\varphi }_{1,1,1}+\,i_c\sin (\varphi _{1,1,1})=I_{1,1}^a-\gamma
\\ \stackrel{.}{\varphi }_{1,1,2}+\,i_c\sin (\varphi
_{1,1,2})=-I_{1,1}^a+I_{1,2}^a-\gamma \\ \stackrel{.}{\varphi }%
_{1,1,3}+\,i_c\sin (\varphi _{1,1,3})=-I_{1,2}^a-\gamma 
\end{array}
\right. , & \left\{ 
\begin{array}{l}
\stackrel{.}{\varphi }_{0,2,1}+\,i_c\sin (\varphi _{0,2,1})=I_{1,1}^a \\ 
\stackrel{.}{\varphi }_{0,2,2}+\,i_c\sin (\varphi _{0,2,2})=I_{1,2}^a \\ 
\stackrel{.}{\;\varphi }_{1,1,1}+\stackrel{.}{\varphi }_{0,2,1}-\stackrel{.}{%
\varphi }_{1,1,2} \\ \qquad - 
\stackrel{.}{\varphi }_{0,1,1}=0 \\ \stackrel{.}{\varphi }_{1,1,2}+\stackrel{%
.}{\varphi }_{0,2,2}-\stackrel{.}{\varphi }_{1,1,3} \\ \qquad -\stackrel{.}{%
\varphi }_{0,1,2}=0 
\end{array}
\right. 
\end{array}
$$
Eliminating the last two equations allows to obtain explicitly the mesh
currents that are now expressed in terms of sine functions of phases:%
$$
\left\{ 
\begin{array}{c}
I_{1,1}^a= 
\frac{\,i_c}{15}(-4\sin (\varphi _{0,1,1})-\sin (\varphi _{0,1,2})+4\sin
(\varphi _{1,1,1})+ \\ -3\sin (\varphi _{1,1,2})-\sin (\varphi
_{1,1,3})+4\sin (\varphi _{0,2,1})-\sin (\varphi _{0,2,2})) \\ 
I_{1,2}^a= 
\frac{\,i_c}{15}(-\sin (\varphi _{0,1,1})-4\sin (\varphi _{0,1,2})+\sin
(\varphi _{1,1,1})+ \\ +\sin (\varphi _{1,1,2})-4\sin (\varphi
_{1,1,3})+\sin (\varphi _{0,2,1})+4\sin (\varphi _{0,2,2})) 
\end{array}
\right. 
$$
After inserting them in Eq(\ref{sol1}), an explicit system of $7$
differential equations (of which only $5$ are independent) results. To give
a synthetic idea of its structure, only two equations are presented here :%
$$
e)\;\left\{ 
\begin{array}{l}
\stackrel{.}{\varphi }_{0,1,1}+i_c\sin (\varphi _{0,1,1})=\frac{i_c}{15}%
(4\sin (\varphi _{0,1,1})+\sin (\varphi _{0,1,2})-4\sin (\varphi _{1,1,1})+
\\ +3\sin (\varphi _{1,1,2})+\sin (\varphi _{1,1,3})-4\sin (\varphi
_{0,2,1})-\sin (\varphi _{0,2,2})) \\ 
\vdots \\ 
\stackrel{.}{\varphi }_{0,2,2}+i_c\sin (\varphi _{0,2,2})=\frac{i_c}{15}%
(-\sin (\varphi _{0,1,1})-4\sin (\varphi _{0,1,2})+\sin (\varphi _{1,1,1})+
\\ +3\sin (\varphi _{1,1,2})-4\sin (\varphi _{1,1,3})+\sin (\varphi
_{0,2,1})+4\sin (\varphi _{0,2,2})) 
\end{array}
\right. 
$$
Remarkably, this system is very different from the previous ones, having no
term linear in $\varphi _{v,r,c}$. It has the same polarization scheme as
the other circuits and the same coefficients as $b)$ in front of the
interaction terms, which are now sine functions of phases instead of beeing
simply phases. It is definitely not obtained as a direct $L\rightarrow 0$
limit from system $b)$.

\section{General derivation of the equations}

One of the two main points of this paper is that a careful derivation of the
equations for the circuits $b)$, $d)$ and $e)$ leads to systems of equations
different from those already existing in the literature. In the previous
example it was shown that a series of manipulations leads to an interaction
that has a range longer than what precedently thought. This point will be
discussed more explicitly in the next section. In Section $5$ it will be
shown that, even though the three circuits seem to be rather different
systems, their equations can be derived from a unique hidden structure and
that it is misleading to consider the limit $L\rightarrow 0$ directly from
equations like those just obtained. To discuss more easily the physics
involved, two steps will be followed. First, the generalization of the
preceding derivation will be illustrated, then a new set of variables will
be discussed. In these variables it will be possible to smoothly handle the
limit $L\rightarrow 0$.

For any $N_r\times N_c$ array the relation 
\begin{equation}
\label{euler}n_m=n_b-n_n+1 
\end{equation}
holds, where $n_m=N_rN_c$ is the number of meshes, $n_b=2N_rN_c+N_r+N_c$ is
the number of branches and $n_n=(N_r+1)(N_c+1)$ is the number of nodes.
Using a matrix notation, the current vector $i=\{i_{v,r,c}\}=\{i_k,k=1,%
\ldots ,n_b\}$, the phase vector $\varphi =\{\varphi _{v,r,c}\}=\{\varphi
_k,k=1,\ldots ,n_b\}$, the mesh current vector $I^a=\{I_{r,c}^a\}=%
\{I_k^a,k=1,\ldots ,n_m\}$, the external bias current vector $%
T_a=\{I_{r,c}\}=\{I_k,k=1,\ldots ,n_m\}$ and the frustration vector $%
f=\{f_{r,c}\}=\{f_k,k=1,\ldots ,n_m\}$ are introduced . In current and phase
vectors the elements are progressively ordered by an index $k$ obtained from
the indices of their elements as $k=(2N_c+1)(r-1)+vN_c+c$, $r=1,\ldots
,N_r+(1-v)$, $c=1,\ldots ,N_r+v$: for example, current $i_{1,1,2}$ of
circuit $a)$ becomes $i_4$. In words, starting from the bottom left hand
corner and following the row of nodes, first come the horizontal branches
then come the vertical branches, and then next node row is scanned. In
frustration and external current vectors the elements are progressively
ordered by an index $k=N_c(r-1)+c$, where $r=1\ldots ,N_r$ and $c=1,\ldots
,N_c$, and the same order is assigned to the meshes, considered as rings
clockwise oriented. Two matrices $A_a$ and $M_a$ are now built \cite{deso}.
The matrix $A_a$, with $n_n$ rows and $n_b$ columns, describes the Kirchhoff
law for current conservation at the nodes. Its element $(A_a)_{m,n}$ is $1$
if in the $m$-th node the $n$-th current enters the node, it is $-1$ if the
current leaves it, it is $0$ otherwise. The matrix $M_a$, with $n_m+1$ rows
and $n_b$ columns, describes the sums around meshes. Its element $%
(M_a)_{m,n} $ is $1$ if in the $m$-th mesh the $n$-th current belongs to the
mesh and is oriented in the same direction of the mesh, it is $-1$ if the
current is oriented in the opposite direction and $0$ otherwise. The matrix $%
M_a$ has one row more than $n_m$ because it includes the so called external
mesh, which follows the rule that $(M_a)_{n_b+1,n}$ is $-1$ if the $n$-th
current is on the boundary of the array and is oriented clockwise following
the boundary ring, $1$ if it is oriented in the opposite direction and $0$
otherwise. It can be proved using network theory \cite{deso} that rows of $%
A_a$ and $M_a$ are mutually ortogonal, and this is also seen by direct
inspection. Since rank$(A_a)=$ $n_n-1$ and rank$(M_a)=$ $n_m$, from Eq.(\ref
{euler}) rank$(A_a)+$rank$(M_a)=n_b$ follows, which means that $A_a$ and $%
M_a $ divide the $n_b$ dimensional space in two parts. Deleting a row in
each of the two matrices does not change their rank: as a convention the
last row will be deleted in $A_m$ to form a new matrix $A$, and the last row
will be deleted in $M_m$ to form a new matrix $M$. These two matrices have
maximum rank and the property 
\begin{equation}
\label{prop}\left\{ 
\begin{array}{l}
MA^T=0,\;\;AM^T=0 \\ 
M^T(MM^T)^{-1}M+A^T(AA^T)^{-1}A=1 
\end{array}
\right. 
\end{equation}
having $AA^T$ and $MM^T$ non-zero determinant. They appear to be dual in the
sense of linear algebra and the quantities 
\begin{equation}
\left\{ 
\begin{array}{l}
M^T(MM^T)^{-1}M=K \\ 
A^T(AA^T)^{-1}A=\overline{K} 
\end{array}
\right. 
\end{equation}
can be seen as projectors on the $n_b$ dimensional space. The induced flux
vector can be introduced as 
\begin{equation}
\label{selfind}\phi _{r,c}=M^Li 
\end{equation}
where the {\it self-induced flux matrix} $M^L$ expresses a relation
analogous to Eq.(\ref{indu1}) or Eq.(\ref{indu2}). This matrix is built from 
$M$ substituting in $M$ for every non-zero entry the {\it branch inductance} 
$\lambda _{m,n}$ if the entry is $1$ and $-\lambda _{m,n}$ if the entry is $%
-1$. Eq.(\ref{indu1}) is recovered when all $\lambda _{m,n}$ are equal to $L$%
, that is $M^L=LM$. In the general case $M^LA^T\neq 0$ and $A(M^L)^T\neq 0$.
The last element is deleted from vector $T_a$ to form vector $T$. Since $T_a$
elements obey current conservation law $\sum_kI_k=0$, the deleted element is
always obtained as $I_{n_m}=-\sum_{k=1}^{n_m-1}I_k$. In this notation
equations Eq.(\ref{kirch}), Eq.(\ref{fluxoid}) and Eq.(\ref{rsj}) can be
rewritten as 
\begin{equation}
\label{compact}\left\{ 
\begin{array}{l}
Ai=T \\ 
M^Li+M\varphi =f \\ 
\stackrel{.}{\varphi }+i_c\sin {}_k(\varphi )_k=i 
\end{array}
\right. 
\end{equation}
where in $\sin {}_k(\varphi )_k$ the label $k$ is to remind the vectorial
form. It is stressed that notation $F\sin {}_k(G)_k$ indicates that the
generic row $k$ of matrix $G$ appears as the argument of the k-th element of
a vector defined as $\sin {}_k(G)_k=\{\sin {}(G_t),t=1,\ldots ,n_b\}$, on
which matrix $F$ operates: e.g., if $F=\left( 
\begin{array}{cc}
F_{11} & F_{1,2} \\ 
F_{2,1} & F_{2,2} 
\end{array}
\right) $ and $G=\left( 
\begin{array}{c}
G_1 \\ 
G_2 
\end{array}
\right) $ then $F\sin {}_k(G)_k=\left( 
\begin{array}{l}
F_{1,1}\sin G_1+F_{1,2}\sin G_2 \\ 
F_{2,1}\sin G_1+F_{2,2}\sin G_2 
\end{array}
\right) $. Clearly, $k$ is not an index. The equations for circuit $b)$ are
now quickly rederived. The general solution of $Ai=T$ is the sum of the
solution of the homogeneous equation $Ax=0$ and a particular solution of $%
Ax=T$. Using Eq.(\ref{prop}), $i=M^TI^a+A^TD$ where $D=(AA^T)^{-1}T$.
Operating with $M^L$ on both sides of this equation results in $%
M^Li=M^LM^TI^a+M^LA^TD$, Eq.(\ref{mu}), that after inversion gives $%
I^a=(M^LM^T)^{-1}M^Li-(M^LM^T)^{-1}M^LA^TD$, Eq.(\ref{mi}). From the second
equation of system (\ref{compact}) it follows that $M^Li=f-M\varphi $, Eq.(%
\ref{fluxoid}), which inserted in the equation for $I^a$ gives $%
I^a=-(M^LM^T)^{-1}M\varphi +(M^LM^T)^{-1}f-(M^LM^T)^{-1}M^LA^TD$. Finally,
using the Josephson equation and defining 
\begin{equation}
\label{k}\left\{ 
\begin{array}{l}
K=M^T(MM^T)^{-1}M \\ 
K^L=M^T(M^LM^T)^{-1}M \\ 
K^{LL}=M^T(M^LM^T)^{-1}M^L 
\end{array}
\right. 
\end{equation}
the equation 
\begin{equation}
\label{vecchiol}\stackrel{.}{\varphi }+i_c\sin {}_k(\varphi )_k=-K^L\varphi
+M^T(M^LM^T)^{-1}f\;-(K^{LL}-1)A^TD 
\end{equation}
for systems $b)$ and $d)$ is recovered. A comment about the conditions under
which $M^LM^T$ can be inverted is needed. As a general rule $\det
(M^LM^T)\neq 0$ if at least $n_m$ branch inductances are non-zero and every
mesh has at least one branch with non-zero inductance. This is the case of
circuits $b)$ and $d)$ but not of circuit $e)$. Moreover, it can be shown
that for any array dimension no element of $K$ is zero, so that $K$
generates an infinite range interaction among phases.

In the case of circuit $e)$ starting equations are 
\begin{equation}
\label{l0}\left\{ 
\begin{array}{l}
Ai=T \\ 
M\varphi =f \\ 
\stackrel{.}{\varphi }+i_c\sin {}_k(\varphi )_k=i 
\end{array}
\right. 
\end{equation}
If $f=0$ it follows that $M\stackrel{.}{\varphi }=0$. Operating with $M$ on
the Josephson equation gives $M\stackrel{.}{\varphi }=M\,i_c\sin
{}_k(\varphi )_k+Mi$, where the l.h.s. member is zero. Inserting in this
equation the solution $i=M^TI^a+A^TD$ gives $Mi_c\sin {}_k(\varphi
)_k=MM^TI^a$, where Eq.(\ref{prop}) has been used. After inversion, this
results in $I^a=(MM^T)^{-1}Mi_c\sin {}_k(\varphi )_k$. Using again the
Josephson equation $\stackrel{.}{\varphi }+i_c\sin {}_k(\varphi
)_k=M^TI^a+A^TD$ and Eq.(\ref{prop}) the equation for $e)$ quickly follows: 
\begin{equation}
\label{vecchio0}\stackrel{.}{\varphi }+i_c\sin {}_k(\varphi )_k=i_cK\sin
{}_k(\varphi )_k+A^TD 
\end{equation}
If $f\neq 0$ this method doesn't work, because it gives again the same
equation without any trace of $f$. The use of $M\varphi =f$ as a constraint
is critical. When differentiated this equation is always $M\stackrel{.}{%
\varphi }=0$ independently of $f$. This problem can be avoided introducing
another variable $p$ as already done with $I^a$. This variable will be
called {\it cut phase}, in analogy with network theory terminology. More
precisely, the equation $M\varphi =f$ can be solved by the substitution $%
\varphi =A^Tp+M^Ts$ where $s=(MM^T)^{-1}f$. The quantity $M\stackrel{.}{%
\varphi }$ is still zero but at the end of the calculation the result is 
\begin{equation}
A^T\stackrel{.}{p}=(K-1)\,i_c\sin {}_k(A^Tp+M^T(MM^T)^{-1}f)_k+A^TD 
\end{equation}
which can be multiplied by $A$ and inverted to give 
\begin{equation}
\label{peq}\stackrel{.}{p}=-(AA^T)^{-1}A\,\,i_c\sin
{}_k(A^Tp+M^T(MM^T)^{-1}f)_k+(AA^T)^{-1}T 
\end{equation}
where Eq.(\ref{prop}) has been used. After solving this differential system,
phases $\varphi $ are recovered as $\varphi (t)=A^Tp(t)+M^Ts$. This new
variable $p$ seems to better suit to the $L=0$ problem, transforming it in a
system of {\it independent }and {\it explicit} differential equations.
Unlike the case of Eq.(\ref{compact}) it is not possible to eliminate $p$ at
the end of the calculation, as it was did with the intermediate variable $%
I^a $. The reason stands on the fact that system (\ref{l0}) is constrained
by the $n_m$ static equations $M\varphi =f$ and only $n_b-n_m$ variables are
left independent, so that the whole dynamics depends only on these new
variables. Applying this new point of view to the circuit $e)$, the system
of $5$ independent explicit differential equations in the $5$ $p_l$ ($%
l=1,\ldots ,5$) variables 
\begin{equation}
\left( 
\begin{array}{l}
\stackrel{.}{p}_1 \\ \stackrel{.}{p}_2 \\ \stackrel{.}{p}_3 \\ \stackrel{.}{p%
}_4 \\ \stackrel{.}{p}_5 
\end{array}
\right) \negthinspace =\negthinspace \frac{i_c}{15}\negthinspace \left( 
\begin{array}{ccccccc}
9 & 6 & 6 & 3 & 6 & 6 & 9 \\ 
-2 & 7 & 2 & 6 & 7 & 2 & 8 \\ 
-1 & -4 & 1 & 3 & 11 & 1 & 4 \\ 
5 & 5 & -5 & 0 & 5 & 10 & 10 \\ 
1 & 4 & -1 & -3 & 4 & -1 & 11 
\end{array}
\right) \negthinspace \left( 
\begin{array}{c}
\sin (-p_1+p_2) \\ 
\sin (-p_2+p_3) \\ 
\sin (-p_1+p_4) \\ 
\sin (-p_2+p_5) \\ 
\sin (-p_3) \\ 
\sin (-p_4+p_5) \\ 
\sin (-p_5) 
\end{array}
\right) \negthinspace +\negthinspace \left( 
\begin{array}{c}
\gamma \\ 
\gamma \\ 
\gamma \\ 
0 \\ 
0 
\end{array}
\right) 
\end{equation}
results.

This idea suggests a way to write the general system in a form suitable to
take the $L\rightarrow 0$ limit in a smooth way, but before undertaking this
task the difference between Eq.(\ref{vecchiol}) and other existing equations
will be discussed.

\section{Mesh currents and induced fluxes}

It is stated in Eq.(6) from Ref.\cite{maj} that for an array where only the
self-inductance of each lattice mesh is retained, and the mutual inductance
among cells is ignored, the relation between the total flux $\phi _{ij}$ in
one mesh and the mesh current of the same mesh $I_{ij}$ is 
\begin{equation}
\label{their}\phi _{ij}^{tot}=\phi _{ext}-L^{\prime }I_{ij}^a 
\end{equation}
This means that the self-induced flux is proportional to the mesh current.
This equation is used also in Ref.\cite{naka}, \cite{mit} and \cite{bock}
and implicitly in Ref.\cite{fila}. In the present work it has been shown for
hybrid circuit $c)$, which is a particular $1\times 2$ case of the $1\times
\infty $ array of \cite{naka}, that Eq.(\ref{their}) is correct because in
this particular case the induced fluxes are indeed proportional to the mesh
currents. From the second block of Eq.(\ref{mi}) it is seen that Eq.(\ref
{their}) is also valid for the pure circuit $d)$, but it is not in general
valid, e.g. when the magnetic fields induced by the four branch currents
surrounding a cell are all of the same strenght. This happens for example in
circuit $b)$, as it is seen from the first block of Eq.(\ref{mi}). {\it This
is the source of the difference between the equation for circuit }$b)${\it \
derived here and the equations derived using Eq.(\ref{their})}. Yet, there
is a case in which Eq.(\ref{their}) can be used also for circuits similar to 
$b)$ where all the four branches contribute in the same way to the flux. Now
it will be shown that in the formalism developed in this paper it is easy to
work out an interaction in which Eq.(\ref{their}) is anyway valid. To
achieve this goal, the mutual inductance among meshes has to be taken into
account.

The task is to link mesh currents and mutually induced fluxes in such a way
that 
\begin{equation}
\phi _{r,c}=M^{MI}i=M^{MI}M^TI^a=p\,I_{r,c}^a 
\end{equation}
where in analogy with Eq.(\ref{selfind}) $M^{MI}$ is a {\it mutually induced
flux matrix} and $p$ is an adjustable constant. For circuit $b)$ in the
example of Fig.\ref{fig1},

$$
M^{MI}=\left( 
\begin{array}{ccccccc}
-\lambda ^{SI} & -\lambda ^{FN} & \lambda ^{SI} & (-\lambda ^{SI}+\lambda
^{FN}) & -\lambda ^{FN} & \lambda ^{SI} & \lambda ^{FN} \\ 
-\lambda ^{FN} & -\lambda ^{SI} & \lambda ^{FN} & (-\lambda ^{FN}+\lambda
^{SI}) & -\lambda ^{SI} & \lambda ^{FN} & \lambda ^{SI} 
\end{array}
\right) 
$$
where $\lambda ^{SI}$ is a self-inductance and $\lambda ^{FN}$ is a first
neighbour mutual inductance, and%
$$
M=\left( 
\begin{array}{ccccccc}
-1 & 0 & 1 & -1 & 0 & 1 & 0 \\ 
0 & -1 & 0 & 1 & -1 & 0 & 1 
\end{array}
\right) 
$$
Requiring that $M^{MI}M^T=p\,1$, where $1$ is the unitary matrix, gives the
system 
$$
\left\{ 
\begin{array}{l}
4\lambda ^{SI}-\lambda ^{FN}=1 \\ 
4\lambda ^{FN}-\lambda ^{SI}=0 \\ 
4\lambda ^{FN}-\lambda ^{SI}=0 \\ 
4\lambda ^{SI}-\lambda ^{FN}=1 
\end{array}
\right. 
$$
which is solved by $\lambda ^{FN}=\frac{\lambda ^{SI}}4$ and $p=\frac{15}4$.
When such an interaction is taken into account, Eq.(\ref{their}) is
recovered. This model can be called MIMF (Mutually Induced Magnetic Flux)
model. From this result an immediate consideration follows. In a circuit
like $b)$ with only self-inductance taken into account, the interaction
among phases is global: every phase of the system appears in each
differential equation. It is interesting to realize that a local phase
interaction typical of a circuit like $d)$, where in each equation only a
small subset of the phases appears, can be obtained introducing in $b)$ a
global mutual inductance coupling. In this sense interaction range in mutual
inductance among cells and interaction range in coupling among phases are
dual.

\section{The $L\rightarrow 0$ limit}

To focus attention on the structure of the limit, it will be assumed that in
Eq.(\ref{compact}) $M^L=LM$, which means that circuit $b)$ will now be
studied. Equation $M(Li+\varphi )=f$ can be solved by setting $q=Li+\varphi $
where $q=A^Tp+M^Ts$. Equation $Ai=T$ is solved by $i=M^TI^a+A^TD$, and
merging these two relations gives 
\begin{equation}
\varphi =A^Tp-LM^TI^a+M^Ts-LA^TD
\end{equation}
Inserting this definition in the Josephson relation gives 
\begin{equation}
\label{barenew}A^T\stackrel{.}{p}-LM^T\stackrel{.}{I^a}+i_c\sin
{}_k(A^Tp-LM^TI^a+M^Ts-LA^TD)_k=M^TI^a+A^TD
\end{equation}
In a block matrix form, defining the matrices 
\begin{equation}
\begin{array}{ll}
B^T=(A^T,-LM^T), & B^{I^a}=(0,M^T)
\end{array}
\end{equation}
and the vector $c=(p,I^a)$, Eq.(\ref{barenew}) is rewritten as 
\begin{equation}
B^T\stackrel{.}{c}=-i_c\sin {}_k(B^Tc+M^Ts-LA^TD)_k+B^{I^a}c+A^TD
\end{equation}
which after inversion is the differential system 
\begin{equation}
\label{final}\left\{ 
\begin{array}{l}
\stackrel{.}{p}=-\,(AA^T)^{-1}A\,i_c\sin {}_k(A^Tp-LM^TI^a+M^Ts-LA^TD)_k+D
\\ L
\stackrel{.}{I}^a=(MM^T)^{-1}M\,i_c\sin {}_k(A^Tp-LM^TI^a+M^Ts-LA^TD)_k-I^a
\\ D=(AA^T)^{-1}T,\;\;s=(MM^T)^{-1}f
\end{array}
\right. 
\end{equation}
In these variables the limit $L\rightarrow 0$ is easier to handle than in a
form like 
\begin{equation}
\stackrel{.}{\varphi }+\,i_c\sin (\varphi )=\frac 1L\varphi 
\end{equation}
In the first row of Eq.(\ref{final}) the interaction between $p$ and $LI^a$
smoothly vanishes and at $L=0$ Eq.(\ref{peq}) is recovered. In the second
row the dynamics gets smoothly frozen and at $L=0$ a static constraint $%
I^a=(MM^T)^{-1}i_cM$ $\sin {}_k(A^Tp+M^Ts)_k$ is imposed to the space of the
degrees of freedom (d.o.f.). The variables $p$ dynamically decouple from $I^a
$ and $I^a$ depend statically on $p$. Multiplying this equations by $MM^T$
gives $Mi=Mi_c\sin {}_k(A^Tp+M^Ts)_k$ which going back to $\varphi $ is the
condition $M(i-i_c\sin (\varphi ))=0$ that follows from $M\stackrel{.}{%
\varphi }=0$. The $n_b$ dimensional d.o.f. space is separated in two
sectors, the $n_m$ dimensional mesh currents sector and the $n_b-n_m$
dimensional cut phases sector, and dynamics survives only in the cut phases
sector. This is a nice example of what could be called a parametric {\it %
dynamical bifurcation}, in analogy with the usual parametric Hopf-like
bifurcation \cite{tabor} of dynamical system theory. When this differential
system is to be studied by perturbation theory a singular problem is
obtained, but in this form the stiffness is less harmful. When $L$ is very
small but not zero, the time variable $t$ in the derivative $d/dt\;I^a$ can
be replaced by $t/L$ to use a dilated time scale. In this scale the $p$
appear almost frozen. The almost slaved $I^a$ tend to follow the dynamics of
the $p$, yet display a second faster but strongly damped dynamics. Moreover,
the behaviour of Eq.(\ref{final}) under $L\rightarrow 0$ limit is physically
satisfactory, because it is hardly belivable that changing smoothly the
inductance of these circuits, e.g. making them bigger, should give some
sharp effect only and exactly at $L=0$. When $M^L$ instead of $LM$ is used,
notation becomes heavier, but it can be shown that only $I_k^a$ belonging to
meshes with zero inductances around the cell become constraints while the
remaining $I_k^a$ keep on taking part in the dynamics.

Eq.(\ref{final}) hides a nice asymmetry or duality between external currents
and external magnetic fluxes. To make it explicit, for $L\neq 0$ the change
of variables is made 
\begin{equation}
\label{uw}\left\{ 
\begin{array}{l}
u=p-LD \\ 
w=-LI^a+s 
\end{array}
\right. 
\end{equation}
Multiplying the first row of Eq.(\ref{final}) by $A^T$ and the second row by 
$M^T$, and substituting in it Eq.(\ref{uw}), gives 
\begin{equation}
\left\{ 
\begin{array}{l}
A^T 
\stackrel{.}{u}=-\,\overline{K}\,i_c\sin {}_k(A^Tu+LM^Tw)_k+A^TD \\ M^T%
\stackrel{.}{w}=-Ki_c\sin {}_k(A^Tu+LM^Tw)_k-\frac 1LM^Tw+\frac 1LM^Ts 
\end{array}
\right. 
\end{equation}
and taking away again $A^T$ from the first row and $M^T$ from the second row
gives 
\begin{equation}
\label{steps}\left\{ 
\begin{array}{l}
\stackrel{.}{u}=-\,(AA^T)^{-1}A\,i_c\sin {}_k(A^Tu+LM^Tw)_k+D \\ \stackrel{.%
}{w}=-(MM^T)^{-1}Mi_c\sin {}_k(A^Tu+LM^Tw)_k-\frac 1Lw+\frac 1Ls 
\end{array}
\right. 
\end{equation}
In this form it is clear that external currents $D$ act as a bias only for
the phase sector while external magnetic fluxes $s$ act as a bias only for
the mesh current sector. This is consistent with the physics of the system.
Phases can grow indefinitely for suitable current biases and cannot be
stopped by magnetic counterfields. Mesh currents instead are subject to a
confining potential and an external bias cannot let them grow indefinitely,
it can just modify their confined dynamics. In this case, it is seen from
Eq.(\ref{steps}) that only an external magnetic field and not a bias current
can change the position of mesh current minima, which are the flux dance
steps.

\section{Conclusions}

The first conclusion is that in a mesh analysis a Josephson junction array
SIMF model is described by an equation like Eq.(\ref{vecchiol}). For a
circuit like $b)$ the interaction range of phases is infinite so that it may
be not necessary to take into account mutual inductance effects to explain
finer system features like those studied in Ref.\cite{dominguez} or in Ref.%
\cite{mit}. The second conclusion is that the $L\rightarrow 0$ limit of the
SIMF model is actually the XY model and with Eq.(\ref{final}) this limit can
be studied analytically in a way easier than thought before.

\section*{Acknowledgments}

The author would like to thank Prof. P. Sodano for having proposed to him
the problem, Dr. D.P.\ Li for stimulating criticism and Prof. F. Marchesoni
for encouragement.

\end{document}